\documentclass{cernrep}
\begin{document}
\title{$\gamma\gamma$ and $\gamma p$ measurements with forward proton taggers in CMS+TOTEM}
\author{J. Hollar, on behalf of the CMS and TOTEM Collaborations}
\institute{LIP, Lisbon, Portugal}

\begin{abstract}
The CMS+TOTEM Precision Proton Spectrometer operated for the first time in 2016, collecting data during $pp$ collisions 
at $\sqrt{s} = 13$~TeV at the CERN Large Hadron Collider. Procedures for the detector alignment, optics corrections, and 
background estimations were developed, and applied to an analysis of the process $pp \rightarrow p \mu^{+}\mu^{-} p^{(*)}$ with 
dimuon masses larger than 110~GeV. A total of 12 candidate events are observed, corresponding to an excess of $>4\sigma$ over 
the background prediction. This constitutes the first evidence for this process at such masses, and demonstrates the good 
performance of CT-PPS.  
\end{abstract}


\maketitle

\section{Introduction}

The CMS+TOTEM Precision Proton Spectrometer (CT-PPS)~\cite{Albrow:2014lrm} is a joint program of the CMS~\cite{Chatrchyan:2008aa} and TOTEM~\cite{Antchev:2013hya} 
collaborations, to operate near-beam forward proton detectors in high luminosity proton-proton running at the Large Hadron Collider (LHC). The detectors consist of 
silicon tracking and fast timing detectors, installed in Roman Pot stations $\sim$~210-220m from the P5 interaction region. The detectors are designed to detect the
intact protons from ``exclusive'' production ($pp \rightarrow pXp$), primarily via either $\gamma\gamma$ fusion, or gluon-gluon interactions (with a second screening 
gluon exchanged to cancel the color flow).

At the LHC, a special class of collisions involves the exchange of quasi-real photons, with the incident beam particles remaining intact. In 
high energy proton-proton collisions, the spectrum of these $\gamma\gamma$ interactions can extend to the TeV scale, well beyond the range probed at 
previous colliders. This provides a unique opportunity to study photon interactions in a new energy regime. Detection of the outgoing protons provides strong 
background suppression and kinematic constraints, making this topology an excellent way to search for new particles and deviations from the Standard Model in 
low cross section processes (for some recent examples see~\cite{deFavereaudeJeneret:2009db, N.Cartiglia:2015gve, Fichet:2014uka, Fichet:2013gsa, Fichet:2016pvq}).  

\subsection{The CT-PPS detectors and 2016 operations}

The initial design of CT-PPS called for the installation and commissioning of detectors in 2016, with physics data-taking to start in 2017. 
However, by making use of the existing TOTEM silicon strip detectors in RP stations at 206m and 214m from CMS, it was possible to begin 
data taking for physics already in 2016. This required validation of safe Roman Pot insertions at high beam intensities, as well as 
integration of the data acquisition and reconstruction software between CMS and TOTEM. This was accomplished at the beginning of the 2016 LHC 
run, and from May the Roman Pots were regularly inserted to 15$\sigma$ from the beam, with the Si-strip detectors read out through the central CMS 
DAQ. During the summer of 2016 diamond fast timing detectors were installed in new cylindrical Roman Pots in the 220m region, and began taking data 
by the end of the LHC proton-proton run. 

\section{Alignment and optics corrections}

\subsection{Alignment}

The alignment is performed in two steps~\cite{Kaspar:2256296}. 
First, special low-luminosity fills are used to determine the absolute alignment. This 
is achieved by inserting the RPs to within $5\sigma$ of the beam. By 
also inserting vertical RPs, a sample of elastic scattering 
$pp \rightarrow pp$ events can be collected, allowing alignment with 
respect to the beam based on the azimuthal symmetry of this process. 
The horizontal RPs can then be aligned with respect to the vertical RPs, 
using a subsample of tracks passing through both. 

\begin{figure}[ht]
\begin{center}
\includegraphics[width=9cm]{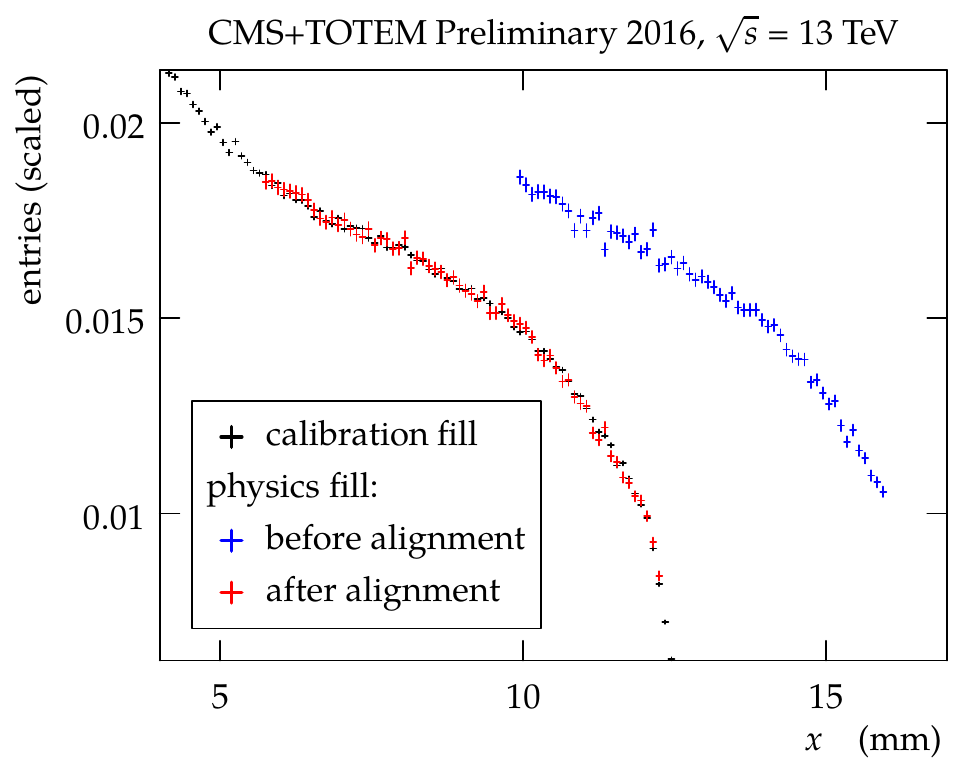}
\caption{Distribution of the track impact points as a function of the horizontal coordinate for the alignment fill (black points), a physics fill before 
alignment (blue points), and after alignment (red points).}
\label{fig:alignment_plot}
\end{center}
\end{figure}

In a second step, the absolute alignment determined in the special fills 
must be transferred to the case of normal high-intensity fills, where 
only the horizontal RPs are inserted to a distance $\sim$15$\sigma$ from 
the beam. This is based on matching the measured x distribution of track 
impact points, under the assumption that the same physics processes contribute 
to all fills. This results in a set of fill by fill alignment corrections, 
determined separately for each RP. The total uncertainty of the horizontal 
alignment procedure is on the order of 150 $\mu$m. The procedure is illustrated in 
Fig.~\ref{fig:alignment_plot}.

\subsection{Optics}

Having aligned the RPs, a precise determination of the LHC beam optics 
is required to derive the proton's fractional energy loss $\xi$ from the 
measured x coordinates of the tracks. The procedure developed~\cite{Nemes:2256433} relies on using real data to 
constrain the elements of the single pass transport matrix $T(s,\xi)$, whose elements are the optical 
functions of the beam line. 

\begin{figure}[ht]
\vspace*{-40mm}
\begin{center}
\includegraphics[width=0.7\textwidth]{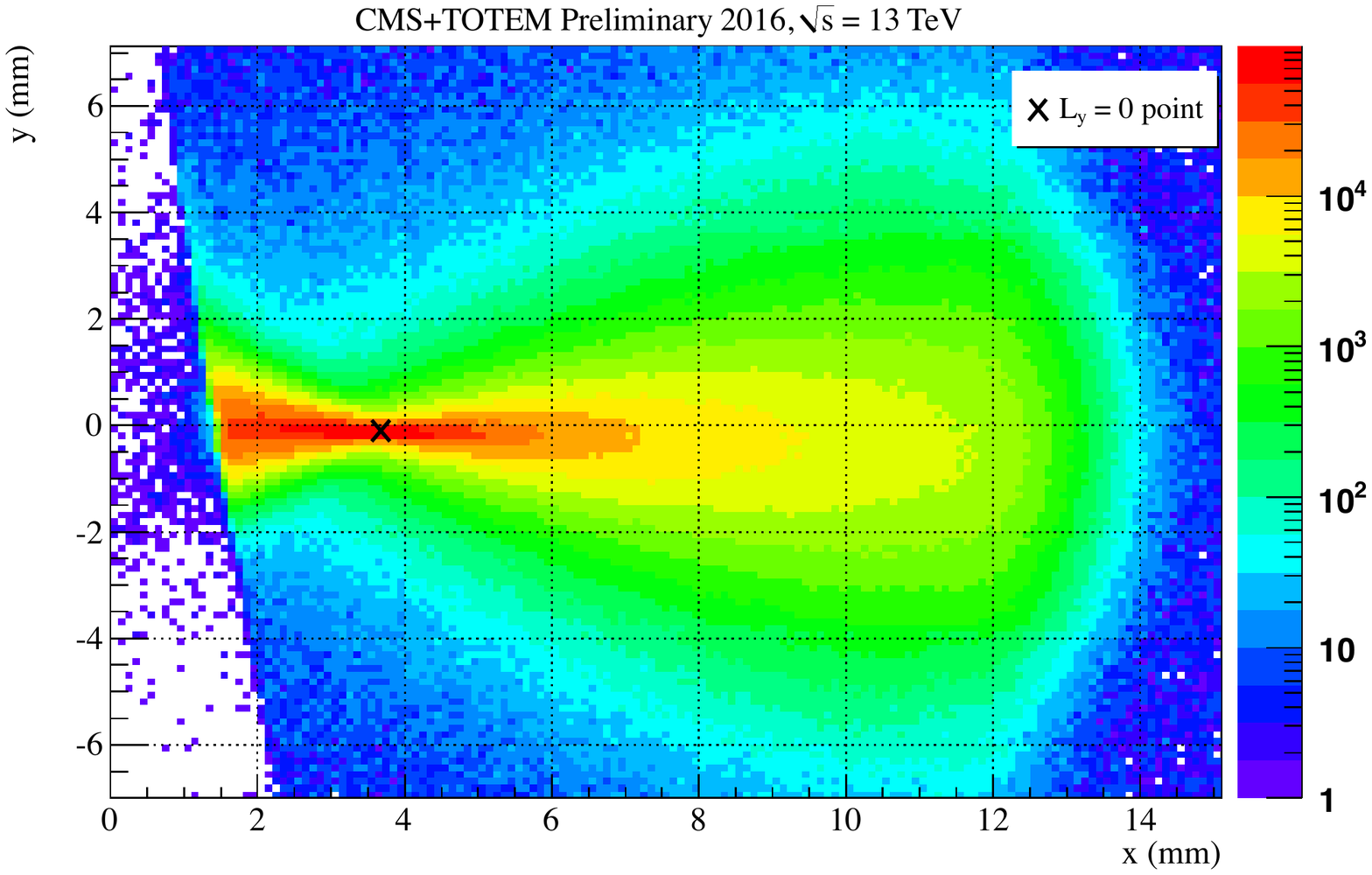}
\vspace*{-40mm}
\caption{Distribution of the track impact points measured in RP 210F, in sector 45, in the alignment fill.
The point where $L_{y}=0$  and its effect in the impact point distribution
are shown. The beam center is at $x=y=0$.}
\label{fig:Ly_plot}
\end{center}
\end{figure}

The leading horizontal term in the transport matrix is x = $D_{x}$($\xi$) $\xi$, where $D_{x}$ is the 
dispersion, which has a mild dependence on $\xi$. The leading term in the vertical plane 
is $y\approx L_{y}(\xi)\Theta_{y}^{*},$ where $L_{y}(\xi)$ is the vertical effective length and 
$\Theta_{y}^{*}$ is the vertical angle of the proton at the interaction point. The value of $ L_{y}(\xi)$ will 
go to zero at a particular value of $\xi_0$, leading to a ``pinch'' in the vertical distribution of tracks 
reconstructed in the RP. By determining the horizontal position of this ``pinch'', the value of the 
dispersion can be solved for as $x_{0}\approx D_{x} \xi_{0}\,$ (Fig.~\ref{fig:Ly_plot}) where higher-order terms are 
neglected and included in the systematic uncertainties. 

A second independent method is also used to determine the difference in the dispersions in the two LHC beams, by comparing the measured physics 
proton distribution in the RPs.

\subsection{Uncertainties}

Given the alignment and optics corrections described in the previous sections, the proton $\xi$ can be reconstructed from the 
measured x position of the tracks in the strip detectors of the horizontal RPs. For large values of $\xi$ the dominant uncertainty 
in this determination is $5.5\%$, arising from the dispersion $D_x$. Additional uncertainties come from the alignment ($\sigma(x) \approx 150 \mu$m), 
and the neglected terms in the horizontal terms of the transport matrix ($\sigma(x) \approx 100 \mu$m).

\section{Analysis of $\gamma\gamma \rightarrow \mu^{+}\mu^{-}$ production}

The alignment and optics procedures are then applied to an analysis of $\gamma\gamma \rightarrow \mu^{+}\mu^{-}$ production~\cite{CMS:2017uey}, using 
$10$~fb$^{-1}$ of data collected during 2016. In order to increase the acceptance at lower masses, only one detected proton is required. 
The signal therefore contains a mix of both $pp \rightarrow p\mu^{+}\mu^{-}p$ events, and $pp \rightarrow p\mu^{+}\mu^{-}p^{*}$ events, 
in which one of the protons dissociates into an undetected system $p^{*}$. 

\subsection{Event selection and proton-dimuon matching}

An event sample enriched in $\gamma\gamma$ interactions is selected in the central CMS detectors following a 
procedure similar to that used in earlier studies, in which no detection of the protons was possible. 
Events are required to have a dimuon vertex with no additional tracks within a veto region of $0.5$~mm. The muons 
are required to have a transverse momentum $p_{T} > 40$~GeV, and invariant mass $m(\mu\mu) > 110$~GeV. In addition, 
the ``acoplanarity'' ($1 - |\Delta \phi(\mu\mu)|/\pi$) of the muons is required to be less than 0.009. The selection 
criteria are chosen such that the expected signal to background ratio after the central detector requirements is $> 1$. 
Because of the high rate of multiple collisions within the same bunch crossing (``pileup''), the selection is based on 
information from reconstructed charged tracks and muons, without using information from the calorimeters. 

In the case of events in which both protons stay intact ($pp \rightarrow p\mu^{+}\mu^{-}p$), the kinematics of the muons and the 
protons can be precisely related via the expression:

$$\xi(\mu\mu) = \frac{1}{\sqrt{s}} (p_{T}(\mu_{1}) e^{\pm \eta(\mu_{1})} + p_{T}(\mu_{2}) e^{\pm \eta(\mu_{2})}).$$

When only one of the two protons remains intact ($pp \rightarrow p\mu^{+}\mu^{-}p^*$), the same expression approximately holds when the 
mass of the dissociating system $M_X$ is small; the deviation becomes comparable to the experimental dimuon resolution only for masses 
$M_X \geq 400$~GeV, corresponding to a small fraction of events surviving the zero extra tracks requirement. The signal region is defined 
to include events where $\xi(\mu\mu)$ and $\xi$(RP) match within 2$\sigma$ of their combined experimental resolution. 

\subsection{Backgrounds and systematics}

After the central CMS detector selection, the dominant backgrounds are expected to arise from 
Drell-Yan $\mu^{+}\mu^{-}$ production, and from $\gamma\gamma \rightarrow \mu^{+}\mu^{-}$ production 
with both protons dissociating. These processes can mimic the signal when they occur in the same bunch 
crossing as a proton from a pileup collision, or a Roman Pot track arising from beam-related 
backgrounds. A control sample of $Z \rightarrow \mu^{+}\mu^{-}$ events is used to estimate the probability of 
a high mass $\mu^{+}\mu^{-}$ event overlapping with an uncorrelated RP track. The $\xi(\mu^{+}\mu^{-})$ distribution in the 
control region is reweighted to match the distribution predicted by the Drell-Yan simulation for events entering the signal 
region. Simulation is then used to extrapolate to the number of such events passing the central detector selection on the 
track multiplicity and acoplanarity. In the case of double dissociation backgrounds, simulated events are normalized to the 
predicted number passing the central detector selection from simulation, and randomly mixed with protons from the 
$Z \rightarrow \mu^{+}\mu^{-}$ data control sample. 

Systematic uncertainties on the background yield include those arising from the statistical uncertainty in the $Z \rightarrow \mu^{+}\mu^{-}$ 
sample used to estimate the backgrounds. For the Drell-Yan background, the modeling of the track multiplicity in the simulation, and the effect 
of reweighting the $\xi(\mu^{+}\mu^{-})$ distribution, are also considered as systematic uncertainties. For the double dissociation background, 
uncertainties in the integrated luminosity, and in the theoretical predictions of the survival probability~\cite{Harland-Lang:2016apc}, are included as systematic uncertainties. 
The dominant uncertainties are due to the effect of reweighting the $\xi(\mu^{+}\mu^{-})$ distribution ($25\%$, taken as the full difference between 
the results with and without reweighting), and the modeling of the track multiplicity distribution ($28\%$, taken as the full difference between 
data and simulation in the region with 1-5 extra tracks at the dimuon vertex).

The event-by-event uncertainty on $\xi(\mu\mu)$ is estimated to be $3.3\%$, based on simulation and data-simulation corrections derived 
from $Z \rightarrow \mu^{+}\mu^{-}$ events. The uncertainty on $\xi$(RP) is taken to be $5.5\%$, as described in section 2. 

The total background estimate, including systematic uncertainties, is $1.47 \pm 0.06$ (stat.) $ \pm 0.52$ (syst.) events, dominated by 
the Drell-Yan backgrounds.  

\subsection{Results}

\begin{figure}[ht]
\begin{center}
\includegraphics[width=6cm]{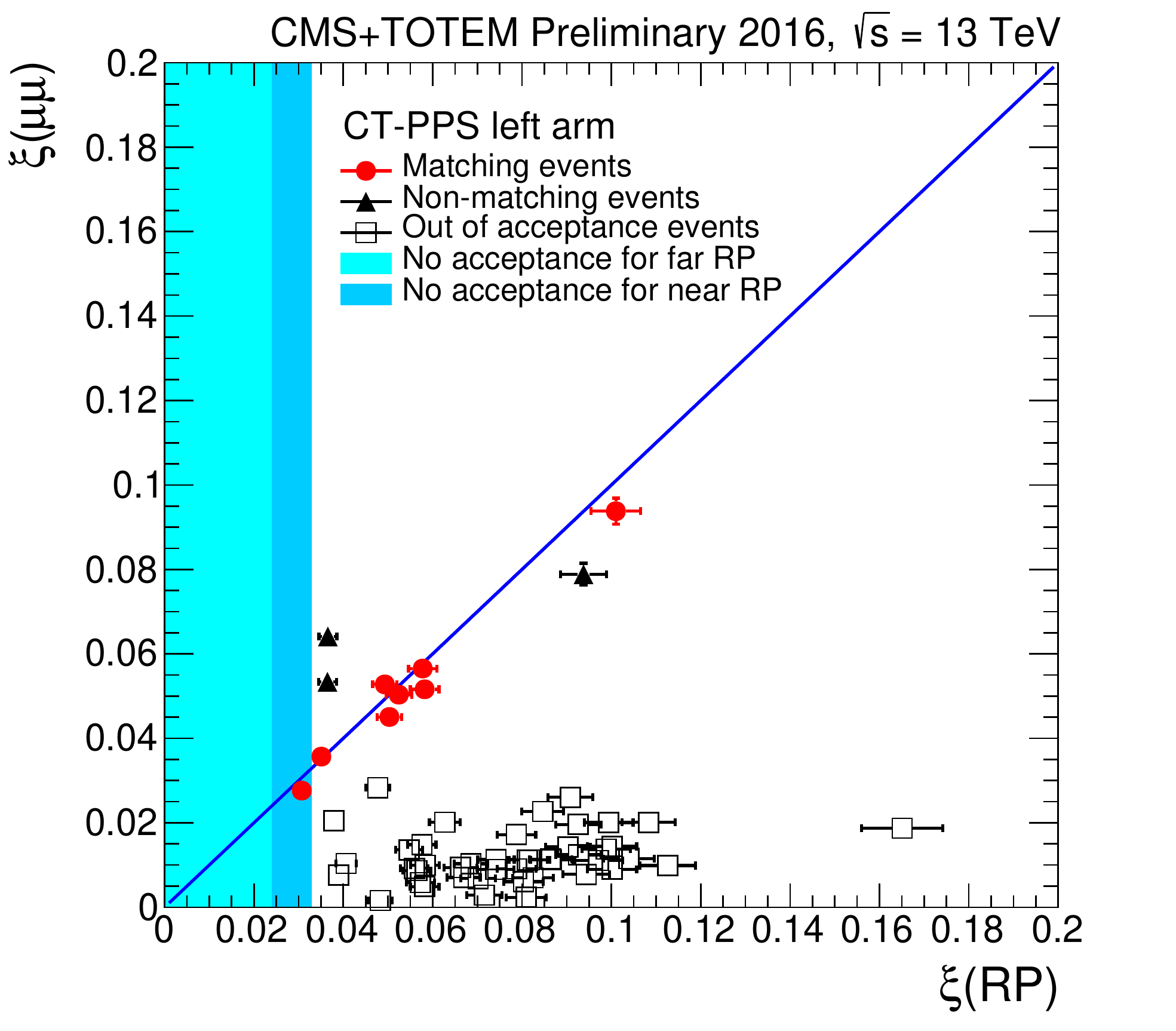}
\includegraphics[width=6cm]{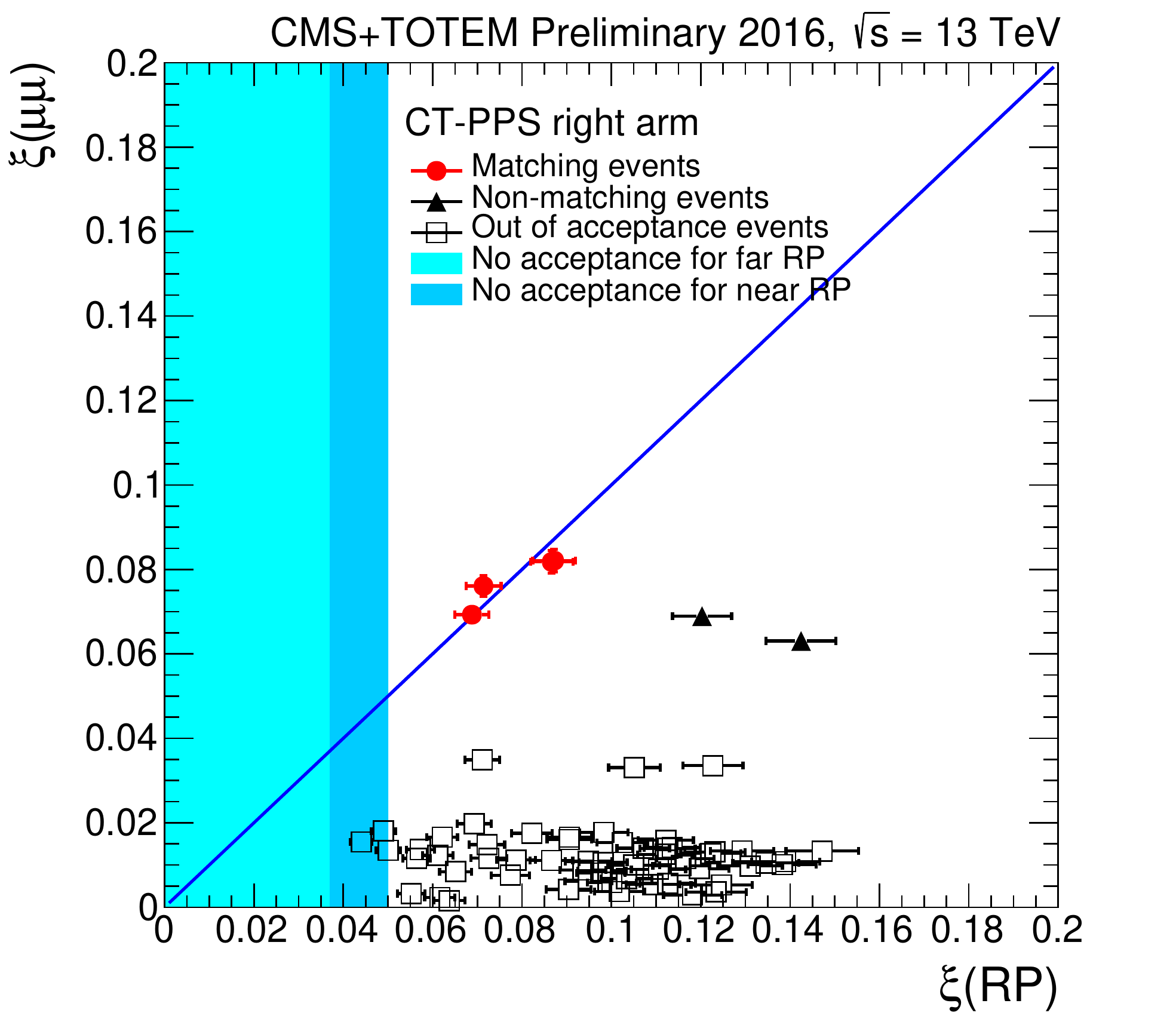}
\caption{Correlation between $\xi(\mu\mu)$ and $\xi$ measured in the Roman Pots, for both Roman Pots in each arm combined. 
The 45 (left) and 56 (right) arms are shown. The light shaded region corresponds to the kinematic region outside the acceptance 
of both the near and far RPs, while the darker shaded region corresponds to the region outside the acceptance of the near RP. For the events in which 
a track is detected in both, the $\xi$ value measured at the near RP is plotted.}
\label{fig:corr_plot_combined}
\end{center}
\end{figure}

The correlation between the predicted $\xi(\mu\mu)$ compared and the 
measured $\xi$(RP) is shown in Fig.~\ref{fig:corr_plot_combined}, for 
the two arms separately. In the region of low $\xi(\mu\mu)$, any 
signal protons should be outside of the RP acceptance, and only random 
background correlations are expected. In the region compatible with the 
RP acceptance, 17 events are observed. Of these, 12 have $\xi(\mu\mu)$ and 
$\xi$(RP) compatible within $2\sigma$ of the resolution, compared to 
$1.47 \pm 0.06$ (stat.) $\pm 0.52$ (syst.) such events expected from the 
backgrounds only. The significance for observing 12 matching events, 
including systematic uncertainties, is estimated to be 4.3$\sigma$.

\begin{figure}[ht]
\begin{center}
\includegraphics[width=8cm]{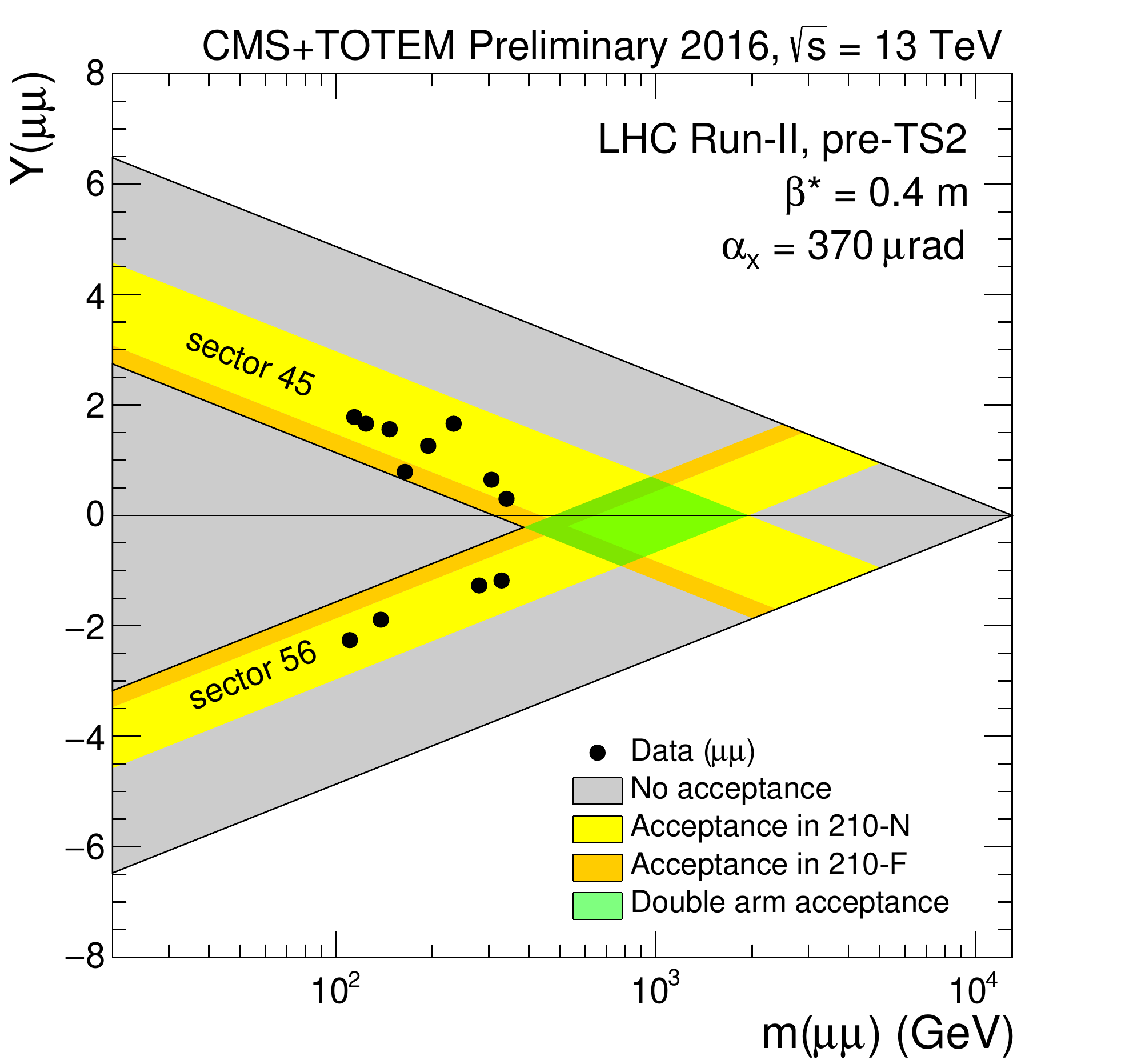}
\caption{Expected approximate coverage in the rapidity vs invariant mass plane, overlaid with the observed dimuon signal candidate events.
}
\label{fig:massrapiditydata}
\end{center}
\end{figure}

In Fig.~\ref{fig:massrapiditydata}, the signal candidate events are overlaid 
with an approximate CT-PPS acceptance (including the assumption that the 4-momentum transfer squared $t = 0$) 
in the dimuon mass-rapidity plane. The events are consistent with the acceptance for detecting 
one of the two protons in CT-PPS. No events with two protons are seen in the data. The 
highest mass candidate has m($\mu\mu$) = 341~GeV, below the region of 
acceptance for detecting both protons.

\section{Conclusions}

The CT-PPS project has a broad physics program, including exploration of new physics in $\gamma\gamma$ interactions at very high energies. 
The detectors operated for the first time in the 2016 LHC proton-proton run, collecting $\sim$~15 fb$^{-1}$ of data. The Roman Pot insertions were 
validated, the detectors were commissioned, and the data acquisition and reconstruction software were fully integrated between CMS and TOTEM. 

Data-driven procedures for the alignment and optics corrections were developed using a combination of special alignment runs, and standard high luminosity 
data taking. These were applied to an analysis of $\gamma\gamma \rightarrow \mu^{+}\mu^{-}$ production with single proton tags. A 4.3$\sigma$ excess of 
events with correlated proton and $\mu^{+}\mu^{-}$ kinematics was observed with m($\mu^{+}\mu^{-}$) = 110-341~GeV, representing evidence for 
tagged $\gamma\gamma$ collisions at the electroweak scale. The present data demonstrate the excellent performance of CT-PPS and its potential. With its 
2016 operation, CT-PPS has proven for the first time the feasibility of operating a near-beam proton spectrometer at a high luminosity hadron collider 
on a regular basis.


%


%
%
%
%
%

\end{document}